\documentclass[aps,prl,twocolumn,notitlepage,superscriptaddress,showkeys,nofootinbib,floatfix,10pt,longbibliography]{revtex4-1}
\usepackage{graphicx,amsmath,verbatim,color,units,nicefrac}
\usepackage[paperwidth=210mm,paperheight=297mm,left=1.8cm,right=1.8cm,top=2.7cm,bottom=2.7cm]{geometry}
\usepackage[charter]{mathdesign}
\definecolor{darkblue}{RGB}{46,48,146}
\usepackage{ifpdf}
\ifpdf
\usepackage[
	pdfstartview=FitH,
	pdfpagemode=UseNone,
  colorlinks=true,
	linkcolor=darkblue,
	filecolor=darkblue,
	urlcolor=darkblue,
	citecolor=darkblue,
	bookmarks=false,
	pdftex,
	plainpages=false,
	hypertexnames=false,
	pdfpagelabels=true,
	hyperindex=true]{hyperref}
\DeclareGraphicsExtensions{.pdf}
\else
\DeclareGraphicsExtensions{.eps}
\fi

\begin{document}

\title{Instrumentation for high-resolution laser spectroscopy at the ALTO radioactive-beam facility}

\author{\mbox{D.~T.~Yordanov}}
\email[]{Deyan.Yordanov@cern.ch}
\affiliation{Institut de Physique Nucl\'eaire, CNRS-IN2P3, Universit\'e Paris-Sud, Universit\'e Paris-Saclay, Orsay, France}

\author{\mbox{D.~Atanasov}}
\affiliation{Experimental Physics Department, CERN, Geneva, Switzerland}

\author{\mbox{M.~L.~Bissell}}
\affiliation{School of Physics and Astronomy, The University of Manchester, Manchester, United Kingdom}

\author{\mbox{S.~Franchoo}}
\affiliation{Institut de Physique Nucl\'eaire, CNRS-IN2P3, Universit\'e Paris-Sud, Universit\'e Paris-Saclay, Orsay, France}

\author{\mbox{G.~Georgiev}}
\affiliation{CSNSM, CNRS-IN2P3, Universit\'e Paris-Sud, Universit\'e Paris-Saclay, Orsay, France}

\author{\mbox{A.~Kanellakopoulos}}
\affiliation{Instituut voor Kern- en Stralingsfysica, KU Leuven, Leuven, Belgium}

\author{\mbox{S.~Lechner}}
\affiliation{Experimental Physics Department, CERN, Geneva, Switzerland}
\affiliation{Technische Universit\"at Wien, Wien, Austria}

\author{\mbox{E.~Minaya Ramirez}}
\affiliation{Institut de Physique Nucl\'eaire, CNRS-IN2P3, Universit\'e Paris-Sud, Universit\'e Paris-Saclay, Orsay, France}

\author{\mbox{D.~Nichita}}
\affiliation{ELI-NP, Horia Hulubei National Institute for R\&D in Physics and Nuclear Engineering, Magurele, Romania}
\affiliation{Doctoral School in Engineering and Applications of Lasers and Accelerators, University Polytechnica of Bucharest, Bucharest, Romania}

\author{\mbox{L.~V.~Rodr\'iguez}}
\affiliation{Max-Planck-Institut f\"ur Kernphysik, Heidelberg, Germany}

\author{\mbox{A.~Said}}
\affiliation{Institut de Physique Nucl\'eaire, CNRS-IN2P3, Universit\'e Paris-Sud, Universit\'e Paris-Saclay, Orsay, France}

\date{\today}

\begin{abstract}
\bf
\noindent
Collinear laser spectroscopy is one of the essential tools for nuclear-structure studies. It allows nuclear electromagnetic properties of ground and isomeric states to be extracted with high experimental precision. Radioactive-beam facilities worldwide strive to introduce such capabilities or to improve existing ones. Here we present the implementation of collinear laser spectroscopy at the ALTO research laboratory, along with data from successful off-line commissioning using sodium beam. The instrumental constituents are discussed with emphasis on simple technical solutions and maximized use of standard equipment. Potential future applications are outlined.

\end{abstract}

\maketitle

\section{Introduction}

\noindent
Laser spectroscopy is an experimental technique probing the energy levels of the atomic hyperfine structure induced by the nuclear electromagnetic properties \cite{Kopfermann1958}. Root mean square charge-radii changes, magnetic-dipole and electric-quadrupole moments of a nucleus are the quantities typically assessed. Atomic beams with well-defined energy of the order of tens of keV in combination with narrow-band lasers facilitate high-resolution measurements limited essentially by the lifetime broadening.\footnote{The natural linewidth in angular frequency is inversely proportional to the lifetime of the excited state: $\Gamma=\tau^{-1}$.} The relativistic Doppler effect:
\[\nu=\nu_0 (1-\beta\cos\phi)/\sqrt{1-\beta^2}\]
is used, where $\phi$ is the angle between the propagation direction of the laser radiation and the velocity $\vec{\upsilon}$ of the atomic beam ($\beta=|\vec{\upsilon}|/c$), and takes values of $\phi=0$ and $\phi=2\pi$ radians for collinear and anti-collinear geometry, respectively. The Doppler-shifted frequency $\nu$ is scanned to match the transitions of the hyperfine structure by varying the velocity $|\vec{\upsilon}|$ while the laser frequency $\nu_0$ is kept constant, or vice versa. In most cases, measurements are implemented on atomic states, which require neutralisation \cite{Klose2012} of the singly-ionized beams used for acceleration. Resonant laser excitations are detectable by various means \cite{Neugart2017}. A comprehensive overview of the technique and its existing implementations can be found in dedicated reviews \cite{Blaum2013,Campbell2016}. The following outlines the newly-constructed collinear-laser-spectroscopy setup at the ALTO facility \cite{Ibrahim2007}.


\section{Instrumentation}

\begin{figure*}[t]
	\begin{center}
		\includegraphics[angle=0,width=\linewidth]{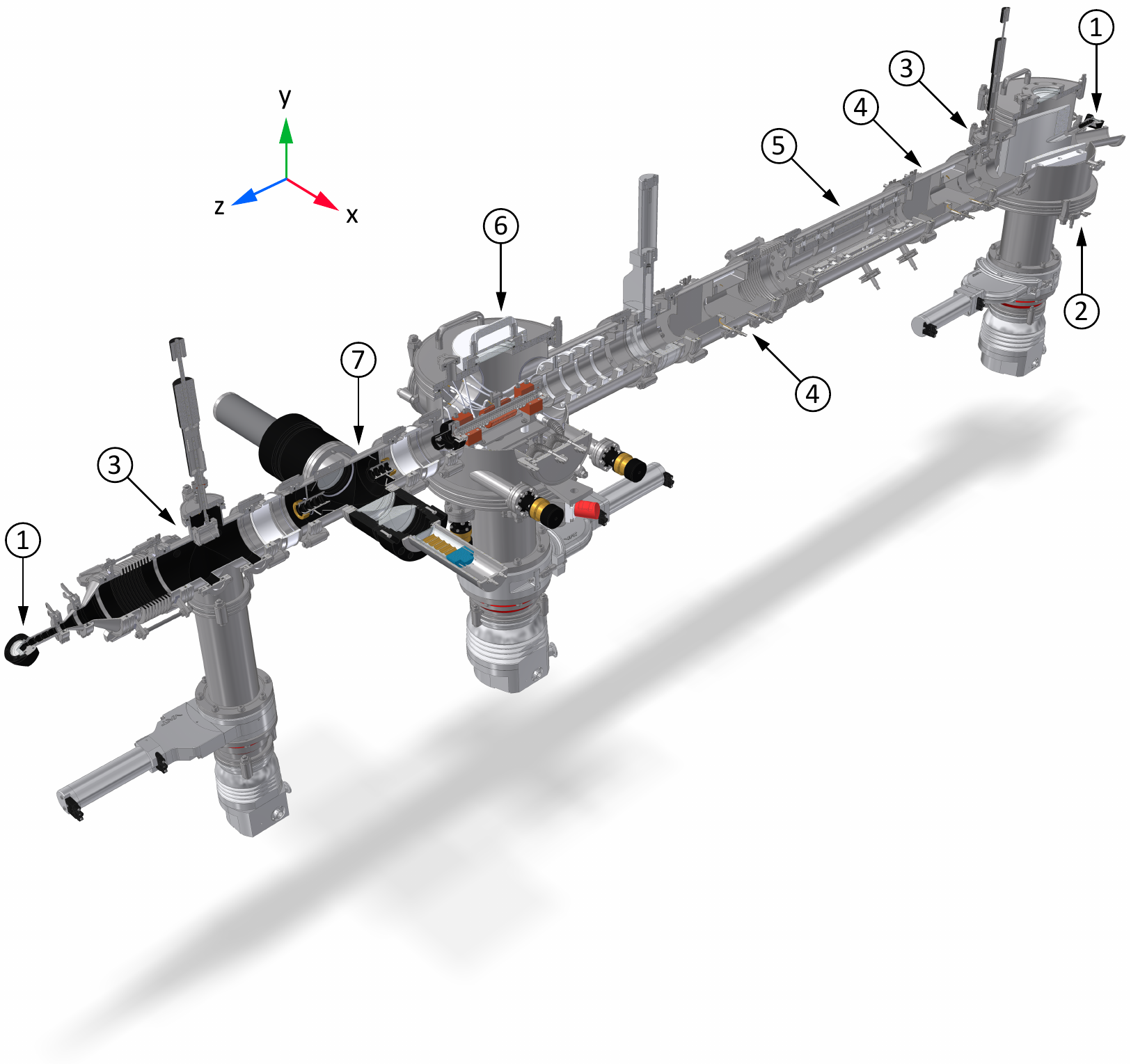}\vspace{-20mm}
			\caption{\textbf{Vacuum beamline.} Technical drawing with a $90^\circ$ longitudinal cross section: 1.)~Fused-silica windows at Brewster's angle; 2.)~$20^\circ$ bender for laser and ion-beam overlap; 3.)~Faraday cups; 4.)~Electrostatic x-y deflectors; 5.)~Electrostatic quadrupole triplet; 6.)~Post-acceleration and charge-exchange chamber; 7.)~Optical detection; Total length: $3.8$~m.}
			\label{fig1}
	\end{center}
\end{figure*}

\noindent
Figure~\ref{fig1} shows a technical drawing of the associated vacuum beamline. Due to space limitations in the experimental hall, the realised assembly is relatively compact, with a total length of $3.8$~m. The instrument is commissioned by measurements in the D$_1$ line of $^{23}$Na, as shown in Fig.~\ref{fig2}. A brief technical description of the instrumentation is presented below.

\begin{figure*}[t]
	\begin{center}
		\includegraphics[angle=0,width=\linewidth]{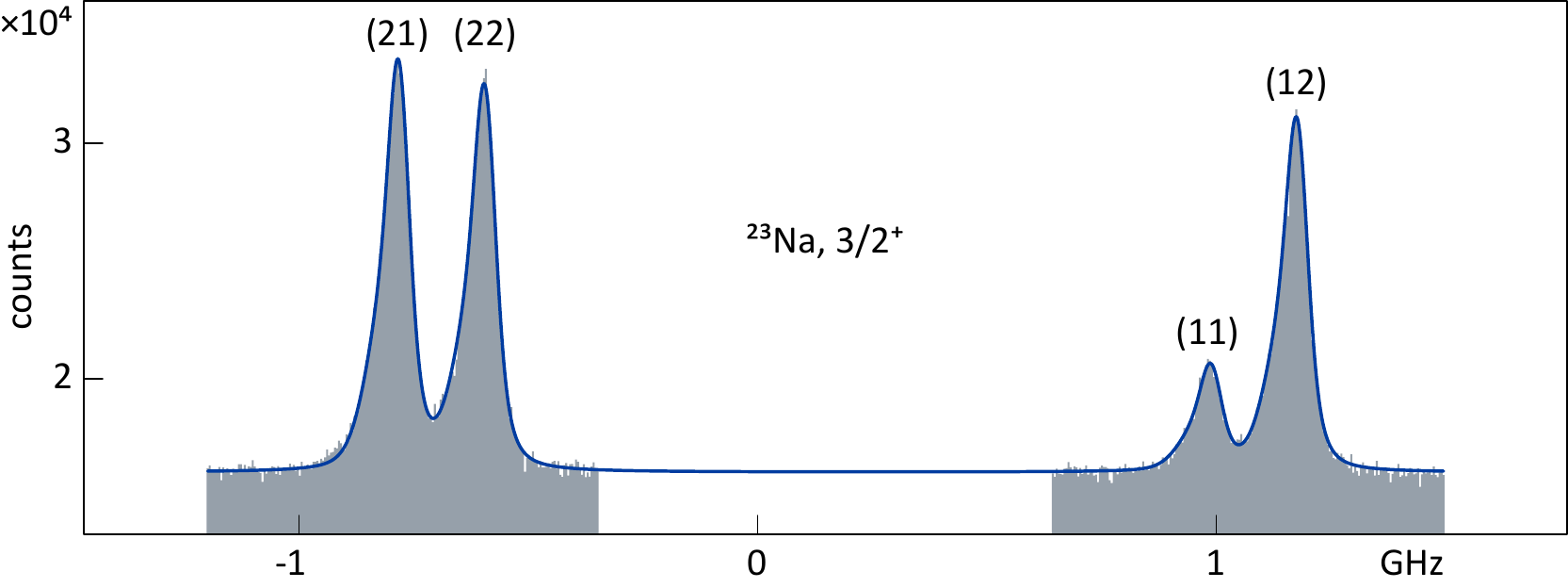}
			\caption{\textbf{Fluorescence spectrum of the D$\boldsymbol{_1}$ line in $\boldsymbol{^{23}}$Na $\boldsymbol{(I^\pi=3/2^+)}$.} Each transition is denoted in parentheses by the total angular-momentum quantum numbers of the lower and the higher state, respectively. The frequency scale is relative to the fine-structure splitting.}
			\label{fig2}
	\end{center}
\end{figure*}

\paragraph{Merging bender} A cylindrical bender with a radius of $800$~mm, a gap of $60$~mm, and a height of $180$~mm facilitates a $20^\circ$ rotation of the ion beam. The laser beam penetrates the outer electrode collinearly to the ion beam through an opening of $20$~mm in diameter. The bending voltage requirement per unit beam energy is $0.15$~V/eV. Windows of fused silica at Brewster's angle of $55.5^\circ$, each including a set of apertures, are installed at both ends of the instrument.

\paragraph{Electrostatic ion optics} The bender is followed by a section for ion-beam handling comprising a Faraday cup, two x-y deflectors, and a non-steering quadrupole triplet in between, namely the model \scalebox{0.8}{EQT 64-15} by \scalebox{0.8}{NEC}. The latter is tuned in conjunction with a quadrupole lens \cite{Perrot2014} in the preceding section of beamline to achieve maximum transmission to the second Faraday cup.

\paragraph{Post-acceleration chamber} Chamber 6 in Fig.~\ref{fig1} is specific to collinear laser spectroscopy and conceptually similar to existing instrumentation \cite{Neugart2017,Klose2012}. A five-cylinder electrostatic lens with elements of $40\times80$~mm in length and diameter, separated by $12$-mm thick insulators, is used for modification of the ion-beam energy. The post-acceleration potential is linearly distributed with a voltage divider. Optionally, the first three elements may also be operated as an Einzel lens to obtain an additional handle on the ion-beam focusing. A hot cell for vaporizing alkali metals is held at the post-acceleration potential. For use with sodium vapour on a sodium ion beam, its central region of $80$~mm in length is heated to about $270^\circ$C to achieve 50\% neutralization. A floating power supply \scalebox{0.8}{FPD-40-008-10} by \scalebox{0.8}{ISEG}, matched to a \scalebox{0.8}{THERMOCOAX 1 NcAc} heating wire, is used for the purpose. A heating and cooling oil circulator \scalebox{0.8}{LAUDA-ECO-RE-630-GN} maintains both ends at a constant temperature of just a couple of degrees above the melting point, hence at a $100^\circ$C for sodium. Apertures of $8$~mm contain the molten metal within the cell while obstructing the propagation of stray light.

\paragraph{Optical detection} Two telescopes of aspheric lenses positioned in the horizontal plane image the ion-beam fluorescence onto the faceplates of photomultiplier tubes. Longitudinal aperture arrays down to $8$~mm at the entrance and $12$~mm at the exit reduce the background from laser scattering. All internal surfaces are painted in colloidal graphite. The lenses \scalebox{0.8}{AL100100-328-SP} by \scalebox{0.8}{THORLABS} with a diameter of $100$~mm are custom made from ultra-violet grade fused silica with a focal length from the flat surface of $76.2$~mm at a wavelength of $328$~nm. The assembly is designed with spacer rings which allows adjusting the focal distance in steps of $5$~mm to accommodate wavelengths in the range from $200$~nm to $1~\mu$m. Head-on photomultiplier tubes for single-photon counting are used, such as the $51$-mm $(2^{\prime\prime})$ model \scalebox{0.8}{9829QSA} from \scalebox{0.8}{ET} with domed quartz windows, magnetic shielding, and a specified dark count rate of less than $300$~Hz.

\paragraph{Control and data acquisition} Analogue and digital voltage input-output is implemented using \scalebox{0.8}{PXI} express electronic modules from \scalebox{0.8}{NATIONAL INSTRUMENTS}. A four-quadrant source and measure unit type \scalebox{0.8}{4137} provides a scanning potential, software limited to $\pm10$~V, with high stability and low ripple. The actual post-acceleration voltage is produced by a $\pm10$-kV four-quadrant fast amplifier \scalebox{0.8}{10/10B-HS-H-CE} from \scalebox{0.8}{TREK} capable of ramping speeds greater than $700~\text{V/}\mu\text{s}$. A digital multimeter type \scalebox{0.8}{4081} is used to monitor and record the applied voltage via an \scalebox{0.8}{OHM-LABS} voltage divider model \scalebox{0.8}{KV-10R}. During a scan, the multimeter is operated in a $\unitfrac[5]{1}{2}$-digit mode at 3000~S/s. An ultra-precise $\unitfrac[7]{1}{2}$-digit mode at 100~S/s is used for independent voltage calibrations. An identical module is available for monitoring the acceleration voltage. The analogue signals of the photomultiplier tubes are converted to low-voltage TTL pulses and read by a timer-counter module type \scalebox{0.8}{6612}, operated as a 32-bit edge counter. The unit is also capable of digital-signal output for controlling auxiliary equipment. All modules are housed in a type \scalebox{0.8}{1082} chassis with an \scalebox{0.8}{8840}-type controller and currently operated with a user-oriented \scalebox{0.8}{LabVIEW} software.


\section{Results of commissioning test}

\noindent
Surface-ionized sodium was accelerated to an energy of $30$~keV and subjected to collinear laser spectroscopy in the transition $3s\;^2S_{1/2}\rightarrow 3p\;^2P^\text{o}_{1/2}$. The required wavelength of $589.8$~nm \cite{Juncar1981} was produced by a cw ring dye laser using Rhodamine 590 in ethylene glycol as the active medium. An example spectrum is shown in Fig.~\ref{fig2} fitted with the empirical lineshape from Ref.~\cite{Yordanov2017}, which accounts for the asymmetric line broadening from collisional excitations in the hot cell after charge exchange. The full width at half maximum of resonances observed throughout the experiment is about $50$~MHz, determined by a substantial Gaussian-like component. Considering the differential Doppler shift of $14.2$~MHz/V, the latter could be attributed to a $10^{-4}$ ripple on the acceleration voltage. For faster transitions, this contribution will be surpassed by the natural linewidth, which in the above transition is only about $10$~MHz \cite{Volz1996}. A magnetic hyperfine parameter of $+885.3(2)(8)$~MHz is found from the individual fit in Fig.~\ref{fig2}. The systematic uncertainty in the second set of parentheses represents the standard deviation of the sample distribution from 26 independent measurements. This result is in line with the value from atomic-beam magnetic resonance \cite{Beckmann1974}. The spectrum in Fig.~\ref{fig2} was obtained in 30 scans of 400 steps, 20~ms each. Laser power of about 1~mW was used on an incident beam of about $300$~pA, measured with the first Faraday cup in Fig.~\ref{fig1}. The corresponding detection efficiency is one photon in $64000$ ions. For faster transitions at the peak of the detector's quantum efficiency of 30\%, compared to 2\% at the above wavelength, the overall detection efficiency is projected to about $1:1000$.

\section{Perspectives}

\noindent
Collinear laser spectroscopy interlinks several cutting-edge technologies: from the production of exotic beams, to lasers, to high-precision measurement instrumentation. In this context, ``off-the-shelf'' laser spectroscopy is out of reach, certainly on the isotope production side where dedicated facilities are required. Commercially available user-oriented laser systems, on the other hand, are already being exploited in most existing installations to enhance operations. Likewise, the instrumentation described here has been developed with an emphasis on simple solutions, thus, maximizing the use of standard equipment to reduce cost and development time, and to improve user accessibility. To exploit its full potential, the apparatus needs to be used in conjunction with ion-beam bunching capabilities \cite{Nieminen2002} for background suppression. Techniques such as laser-induced nuclear orientation \cite{Neugart2008}, collisional ionization \cite{Geithner2008}, and state-selective charge exchange \cite{Vermeeren1992} may also be considered in specific cases. Possible future applications include studies of nuclear structure or facilitating experiments with polarized beams for decay spectroscopy \cite{Hirayama2005} or research on fundamental symmetries \cite{Severijns2011}. The instrument has been constructed along the time-line of the low-energy radioactive beam facility DESIR \cite{Desir2009}, where dedicated techniques for studies of proton-rich nuclei will be developed.

\acknowledgments
This work has been supported by the Orsay Nuclear Physics Institute, the P2IO Laboratory of Excellence, the French National Institute for Nuclear and Particle Physics, the Paris-Sud University, and the European Union's HORIZON~2020 Program under grant agreement no.~654002. We thank the ALTO technical group for their professional assistance.

\paragraph{Author contributions} \mbox{D.T.Y.} initiated the project and contributed in securing financial resources along with \mbox{S.F.}, \mbox{G.G.}, \mbox{E.M.R.}, \mbox{L.V.R.}, and other collaborators. Implementation took place under the technical coordination of \mbox{A.S.} \mbox{M.L.B}, \mbox{A.K.}, \mbox{S.L.}, \mbox{D.N.}, \mbox{L.V.R.}, and \mbox{D.T.Y.} prepared the instrumentation. \mbox{D.A.} developed the software for data acquisition. All authors contributed to the measurements and the preparation of the manuscript.

\paragraph{Data availability} Technical and experimental details are available from the main author upon reasonable request.

\bibliography{bibliography}

\end{document}